\documentclass[12pt]{article}
\usepackage{epsfig}

\def\beq{\begin{equation}}
\def\eeq#1{\label{#1}\end{equation}}
\def\eeqn{\end{equation}}

\def\beqa{\begin{eqnarray}}
\def\eeqa#1{\label{#1}\end{eqnarray}}
\def\eeqan{\end{eqnarray}}

\let\bar=\overbar

\def\Dslash{\not{\hbox{\kern-4pt $D$}}}
\def\dslash{\not{\hbox{\kern-2pt $\del$}}}

\def\msb{{\bar{\ssstyle M \kern -1pt S}}}

\def\Title#1{\begin{center} {\Large {\bf #1} } \end{center}}

\begin{document}

\Title{From Running Gluon Mass to Chiral Symmetry Breaking}

\bigskip\bigskip


\begin{raggedright}

{\it Orlando Oliveira$^{1,2}$\index{Oliveira, O.}, P. Bicudo$^3$, D. Dudal$^4$,
T. Frederico$^2$, W. de Paula$^2$, N. Vandersickel$^4$ \\
$^1$ Departamento de F\'{\i}sica, Universidade de Coimbra, 3004-516 Coimbra, Portugal \\
$^2$ Departamento de F\'\i sica, Instituto Tecnol\'ogico de Aeron\'autica,
        12228-900 S\~ao Jos\'e dos Campos, SP, Brazil \\
$^3$ Departamento de F\'{\i}sica, I.S.T., Av Rovisco Pais, 1049-001 Lisboa, Portugal\\
$^4$ Ghent University, Department of Physics and Astronomy, Krijgslaan 281-S9, B-9000 Gent, Belgium \\
}

\bigskip\bigskip
\end{raggedright}

\section{Introduction}

In recent years the non-perturbative computation of the two point correlation functions of pure Yang-Mills theory
have attracted  a lot of attention.
For what concerns the Landau gauge gluon propagator, the subject of this communication,
lattice QCD simulations, Schwinger-Dyson equations and non-perturbative quantization of Yang-Mills theories
provide essentially the same results. Recent reviews can be found in \cite{Maas2011}. 
In this sense, we can claim to have a fair description of this two point function
over the entire range of momentum. Here we review the large volume lattice simulations performed
by one of the authors, its interpretation in terms of modeling the gluon propagator
and we describe how the infrared propagator can be incorporated into an effective field theory model which approximates
QCD at low energies. This effective theory connects gluon confinement with a gluon mass $m_g$ with
chiral symmetry breaking. Further, the model predicts a particular simple relation between 
 $m_g$ and the light quark masses $m_q$, i.e. that $m^2_q/ m_g$ is constant.
This relation is tested using the solutions of the Schwinger-Dyson equations and found to be valid
 in the low energy regime below the 10\% accuracy level.

From the point of view of lattice simulations there are still some open questions. 
On the lattice, one numerically checks whether the selected Landau gauge configurations
belong to the so-called Gribov region, which is the set of all transverse gauge connections with 
positive Faddeev-Popov operator. Within this set, one still finds Gribov copies, and it is not 100\% 
well established if/how these additional copies influence the propagator in the (very deep) infrared region. 
However, previous simulations suggest that by choosing a different Gribov copy, the accompanying error
lies typically within the statistical error of the propagator - see, for example, \cite{Silva2004}.

\section{The Quenched Lattice Gluon Propagator}

In this section we report on the Landau gauge quenched lattice gluon propagator
\begin{equation}
   D^{ab}_{\mu\nu} (p^2) = \delta^{ab} \left( \delta_{\mu\nu}- \frac{p_\mu p_\nu}{p^2} \right) D(p^2) \,
\end{equation}
computed for large volumes $ L \, a >  3.5$ fm. 
The propagator has been computed previously for huge lattice volumes,
for the SU(3) gauge group up to  $ L \, a \approx 17$ fm \cite{Bogolubsky2009}
 and for SU(2) up to $ L \, a \approx 27$ fm,
but using $a \sim 0.2$ fm \cite{Cucchieri2007}. The non-perturbative physics scale being $\sim 1$ fm,
it is important to check these results by performing simulations at smaller lattice spacings and evaluate
the corresponding finite volume effects. Here we report the computation of the propagator for
 $a = 0.102$ fm ($\beta = 6.0$),  $a = 0.0726$ fm ($\beta = 6.2$) and
 $a = 0.0544$ fm ($\beta = 6.4$) and for various volumes up to $ L \, a \approx 8.2$ fm.
 The results reported in this section are preliminary.

\begin{figure}[t]
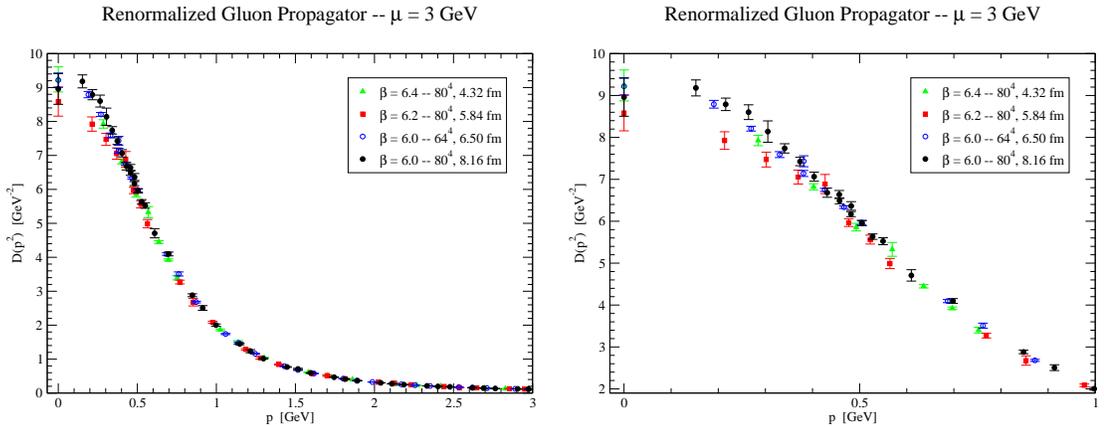

\begin{center}
\epsfig{file=gluonR3GeV.eps,height=2.2in} \,\,
\epsfig{file=gluonIR_R3GeV.eps,height=2.2in}
\caption{$D(p^2)$ for the largest volumes computed for each of the lattice spacings. The plot on the right shows
$D(p^2)$ for the infrared region defined as $p < 1$ GeV.}
\label{fig:gluon}
\end{center}
\end{figure}

In Figure \ref{fig:gluon} we show the renormalized gluon propagator, computed with the different lattice spacings,
for the largest physical volumes up to momenta $p = 3$ GeV and a zoom of the infrared region.
For momenta above 1 GeV all data sets are, within errors, compatible. In the ultraviolet region, i.e. for
$p$ above $\sim 2.8$ GeV, all data sets are well described by the 1-loop inspired fit
\begin{equation}
   D(p^2) = Z \frac{ \left[Ê\ln \frac{p^2}{\Lambda^2} \right]^{-\gamma} }{p^2} \, ,
   \label{Eq:Duv}
\end{equation}
where $\gamma = 13/22$ is the gluon anomalous dimension. Indeed, for each set,
the fits to (\ref{Eq:Duv}) where used to renormalize the gluon propagator according to 
the MOM prescription
\begin{equation}
  \left. D_R (p^2) \right|_{p^2 =  \mu^2} =  \frac{ 1 }{\mu^2} \, .
\end{equation}
The renormalization procedure is described in detail in \cite{Silva2004,Dudal2010}. Note, however, that
in the present work, to renormalize, only the data for momenta above $p \sim 2.8$ GeV was included.
In all the cases the $\chi^2/d.o.f. \sim 1$.

\begin{figure}[t]
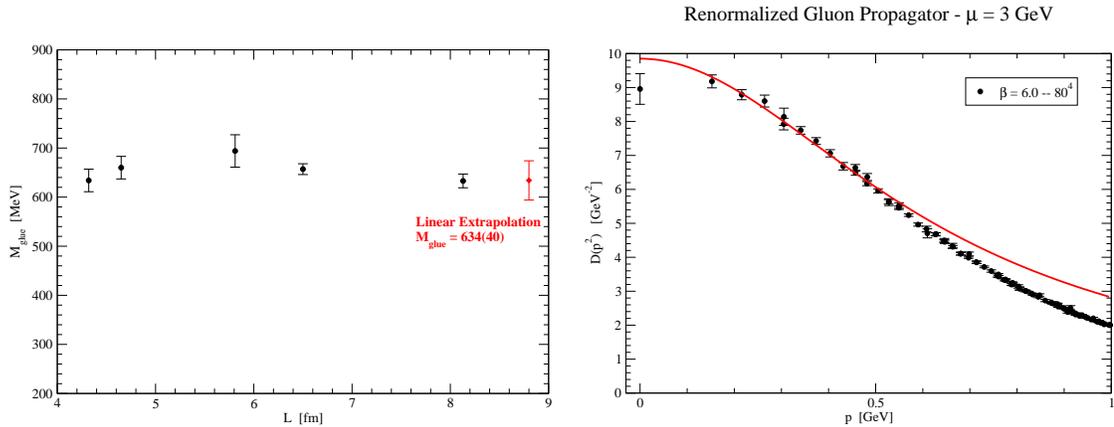

\begin{center}
\epsfig{file=massa_L.eps,height=2.0in} \,\,
\epsfig{file=prop_fit.eps,height=2.2in}
\caption{Gluon mass from the infrared propagator (left) computed assuming a simple pole. The point on the right is
              the extrapolated mass to infinite volume. The right plot illustrates a typical fit.
              Note that the simple pole overestimates $D(0)$.}
\label{fig:Dmass_fit}
\end{center}
\end{figure}

\begin{table}[t]
   \centering
   \begin{tabular}{l@{\hspace{0.8cm}}l@{\hspace{0.8cm}}l} 
   \hline
   $\beta$    &  $L \, a$ & $D(0)$  \\
                   &  (fm) & (GeV$^{-2}$) \\
                      \hline
     6.0      & 6.50 & 9.22(21) \\
     6.0      & 8.16 & 8.96(45) \\
     6.2      & 5.84 & 8.58(43) \\
     6.4      & 4.32 & 9.24(37)\\
        \hline
   \end{tabular}
   \caption{$D(0)$ for the various simulations shown in Figure \ref{fig:gluon}. \label{D0}}
\end{table}

For the infrared region the data sets show finite volume and finite spacing effects. For example, the
propagator computed using the simulation at $\beta = 6.2$, although having a smaller physical volume (5.84 fm)$^4$,
is below all the remaining data sets. For $p = 0$, the large statistical error hides the differences between
the various $D(0)$. For the various simulations reported here, the corresponding $D(0)$ are
given in Table \ref{D0}.
This numbers should be compared with the large volume,$\beta = 5.7$, Wilson action SU(3) simulations 
performed by the Berlin-Moscow-Adelaide (BMA) group. 
Using the same definitions and setting the scale in the same way, i.e. from the string tension, the 
BMA data reads $D(0)$ = 8.68(37), 8.09(36), 7.59(56), 7.17(31) and 7.53(19) GeV$^{-2}$ for
$L \, a$ = 8.09, 11.76, 13.23, 14.70 and 16.17, respectively, given in fm.

Qualitatively, the propagators computed with the different values of $\beta$ are similar.
The $D(p^2)$ from the $\beta = 5.7$ simulations are below all the remaining data sets. 
This suggest that, in order to quote
a continuum propagator, one should extrapolate the lattice propagators to the infinite volume. Recall that, in principle,
the renormalization procedure removes all the lattice spacing dependence. The extrapolation is better achieved if
one uses a theoretical motivated functional form to describe $D(0)$ and extrapolates its parameters to the infinite 
volume - see also \cite{Oliveira2009a,Oliveira2009b}.

Let us assume that the infrared propagator is described by a simple mass pole
\begin{equation}
   D(p^2) = \frac{Z}{p^2 + M^2} \, .
   \label{Eq:Dmass}
\end{equation}
This functional form can only describe the propagator within a limited range of momenta. For example,
a propagator described by (\ref{Eq:Dmass}) does not violate  positivity.
Violation of positivity is a well established property of the non-perturbative gluon propagator.
Anyway, at minimum, a fit to (\ref{Eq:Dmass}) defines an interval range where one can approximate
$D(p^2)$ by such a mass pole. The outcome of the fits as a function of the maximum range of momenta
$p_{max}$ are
\begin{table}[h]
   \centering
   \begin{tabular}{l@{\hspace{0.5cm}}c@{\hspace{0.5cm}}c@{\hspace{0.5cm}}c@{\hspace{0.5cm}}c@{\hspace{0.5cm}}c} 
      \hline
   $\beta$    &  $L \, a$  & $p_{max}$ &  $Z$  &  $M$    & $\chi^2/d.of.$  \\
                   &  (fm)       & (MeV)         &      &  (MeV)     &  \\
                      \hline
     6.0  & 6.50 & 504 & 4.12(10)  &  657(11) & 1.3  \\
     6.0  & 8.16 & 505 & 3.95(12)  &  633(14) & 1.2  \\
     6.2  & 5.84 & 522 & 4.35(30)  &  694(33) & 1.2  \\
     6.4  & 4.32 & 493 & 3.82(20)  &  634(23) &  0.7\\
        \hline
   \end{tabular}
\end{table}

\noindent
It follows that the lattice propagator can be described by a pole mass up to
$p \sim 500$ MeV with a gluon mass between 600 and 700 MeV.
If one assumes that $M(V) = M_\infty + M_1/L$ and fit the lattice data, then the infinite
volume extrapolated mass is $M_\infty = 634(40)$ MeV. The fit has a $\chi^2/d.o.f. = 1.4$.
The data for the different $M$ and the extrapolation can be seen in Figure \ref{fig:Dmass_fit}.
Note that the simple pole (\ref{Eq:Dmass}) overestimates $D(0)$, see right plot in
Figure \ref{fig:Dmass_fit}, and that is the reason why
we do not provide the details of the extrapolation of $Z$ and $D(0)$.

\begin{figure}[t]
\begin{center}
\epsfig{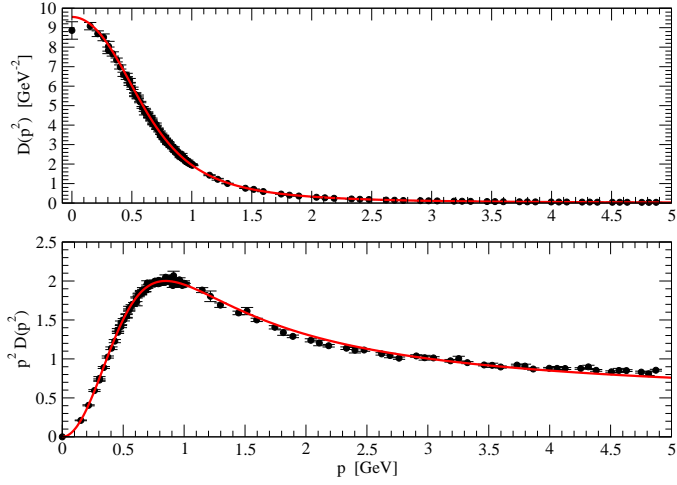}
\caption{Gluon propagator and the fit to (\ref{Eq:Dmass_run}) using the definitions
              (\ref{SDE_M2}). Note that the functional form used here also overestimates $D(0)$.}
\label{fig:D_SDE}
\end{center}
\end{figure}

The lattice data can be described by a propagator of type (\ref{Eq:Dmass}) if $Z$ and $M$ are
functions of momentum, i.e. for
\begin{equation}
   D(p^2) = \frac{Z(p^2)}{p^2 + M^2(p^2)} \, .
   \label{Eq:Dmass_run}
\end{equation}
A momentum dependent gluon mass together with a $Z(p^2)$ were investigated in \cite{Oliveira2011}, where
 the same functional forms were used to fit the decoupling solutions of the Schwinger-Dyson equations.
According to this work the lattice data can be described by
\begin{eqnarray}
  Z(p^2)  =  \frac{z_0}{\left[\ln \frac{p^2 + r \, m^2_0}{\Lambda^2} \right]^\gamma}
  \quad\mbox{ and }\quad
  M^2(p^2)  =  \frac{m^4_0}{p^2 + m^2_0} \label{SDE_M2}
\end{eqnarray}
up to momenta $p_{max} = 4.2$ GeV. The values of the fitted parameters for the largest physical
volume are
\begin{displaymath}
   z_0 = 1.189(20), \quad \Lambda = 1.842(39) \mbox{ GeV},
   \quad r = 7.49(59) \quad \mbox{ and } \quad
   m_0 = 671(9) \mbox{ MeV}.
\end{displaymath}
The gluon data and the fit to (\ref{Eq:Dmass_run}) are reported in Figure \ref{fig:D_SDE}. Note that, as in
the case of simple pole mass, the fit overestimates $D(0)$.

\begin{figure}[t]
\begin{center}
\epsfig{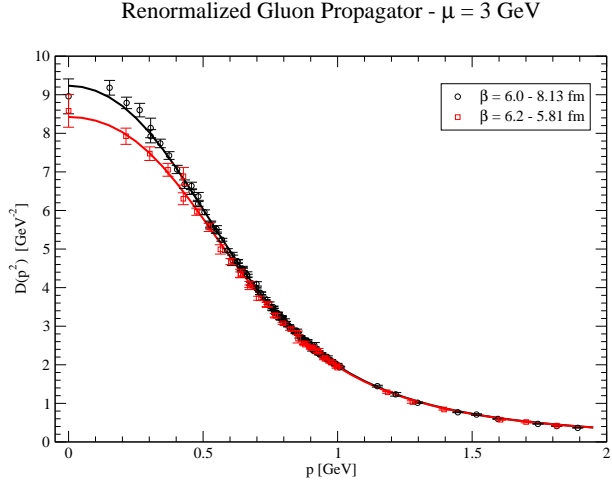}
\caption{Gluon propagator and the fits to (\ref{eq:Drgz}) for the largest lattice volumes. Note that the RGZ reproduces
              well all the lattice data points, including $D(0)$.}
\label{fig:Drgz_fits}
\end{center}
\end{figure}

%
%
\begin{table}[t]
   \centering
   \begin{tabular}{l@{\hspace{0.5cm}}c@{\hspace{0.5cm}}c@{\hspace{0.5cm}}c@{\hspace{0.5cm}}c@{\hspace{0.5cm}}c@{\hspace{0.5cm}}c}
      \hline
     $\beta$      &  $L \, a$  & $\frac{\chi^2}{d.o.f.}$    & $M^2$      &  $M^2 + m^2$  & $\lambda^4$  & $p_{max}$ \\
        \hline
      6.0            &   4.88      & 1.6                      & 2.81(9)    &   0.62(3)            & 0.284(7)         & 1.2   \\
      6.0            &   6.50      & 1.1                      & 2.66(6)    &   0.54(2)            & 0.288(5)         & 0.95 \\
      6.0            &   8.13      & 1.1                      & 2.41(5)    &   0.47(2)            & 0.261(4)         & 1.6 \\
      6.2            &   3.49      & 1.1                      & 2.5(1)      &   0.48(4)            & 0.28(1)           & 1.4 \\
      6.2            &   4.65      & 1.4                      & 2.44(7)    &   0.48(3)            & 0.276(6)         & 1.5 \\
      6.2            &   5.81      & 1.2                      & 2.3(1)      &   0.42(4)            & 0.273(9)          & 1.6\\
         \hline
   \end{tabular}
   \caption{Fits of the lattice data to tree level RGZ propagator. 
                 $La$ is given in fm and the mass scales and higher momenta in the fitting $p_{max}$ are given
                 in power of GeV.}
\label{fitsRGZ}
\end{table}

The refined Gribov-Zwanziger (RGZ) action is an improvement over the usual Faddeev-Popov quantization
procedure for Yang-Mills theories, in the sense that it provides a better way to handle the problem
of the Gribov copies by restricting the functional integration space to the so-called Gribov region. The RGZ
action is renormalizable, in the perturbative sense, and introduces new auxiliary bosonic and fermionic
fields. In what concerns the gluon propagator, the RGZ tree level propagator is given by
\begin{equation}
  D( p^2 ) = \frac{ p^2 + M^2}{p^4 + \left( M^2 + m^2 \right) p^2 + 2 g^2 N \gamma^4 + M^2 m^2} \, ,
  \label{eq:Drgz}
\end{equation}
where $M^2$ is a mass scale related to the new auxiliary fields, $m^2$ is another mass scale related with
the $\langle A^2 \rangle$ condensate and $\gamma^4$ is the Gribov parameter. 
$\gamma^4$ is not a free parameter but is fixed by the so-called horizon condition  \cite{Dudal:2008sp}.
 In the following we
shall introduce the shorthand $\lambda^4 = 2 g^2 N \gamma^4 + M^2 m^2$. The RGZ being a non-perturbative
quantization for the Yang-Mills theories, one hopes that its tree level predictions 
provide a good description for the infrared. The propagator (\ref{eq:Drgz}) can rewritten as
\begin{equation}
  D( p^2 ) = \frac{ 1}{p^2 + M^2(p^2)} \quad\mbox{ with }\quad
  M^2( p^2 ) = m^2 + \frac{ ~ 2 \, g^2 \, N \, \gamma^4 ~}{p^2 + M^2} ~ .
\end{equation}
In this sense, the RGZ action predicts a momentum dependent effective gluon mass which is essentially
the functional form analyzed previously, i.e. $M^2(p^2)$ given by equation (\ref{SDE_M2}). The tree level
expression for $D(p^2)$ does not include the observed logarithmic corrections at high energies and, therefore,
one expects (\ref{eq:Drgz}) to deviate from the lattice data in the ultraviolet region. In order to
extrapolate to the infinite volume and to have an estimate of the error on this extrapolation, we
fit the following two sets of data to (\ref{eq:Drgz}): (i) $48^4$ with $L a = 4.88$ fm, $64^4$ with $L a = 6.50$ fm
and $80^4$ with $L a = 8.13$ fm for $\beta = 6.0$; (ii) $48^4$ with $L a = 3.49$ fm, $64^4$ with $L a = 4.65$ fm
and $80^4$ with $L a = 5.81$ fm for $\beta = 6.2$. The fits are summarized in Table \ref{fitsRGZ}.
The lattice data and the fits for the largest physical volumes are reported in Figure
\ref{fig:Drgz_fits}. Note that the RGZ propagator reproduces well all the lattice data, including $D(0)$.

The infrared propagator can be extrapolated to the infinite volume if one assumes a linear dependence on
$1/(La)$. The extrapolations give
\begin{table}[h]
   \centering
   \begin{tabular}{l@{\hspace{0.9cm}}cc@{\hspace{0.9cm}}cc@{\hspace{0.9cm}}cc}
   \hline
     $\beta$      & $\frac{\chi^2}{d.o.f.}$    & $M^2$ 
                       & $\frac{\chi^2}{d.o.f.}$    & $M^2 + m^2$
                       & $\frac{\chi^2}{d.o.f.}$    & $\lambda^4$  \\
   \hline                       
      6.0            & 1.7 & 1.80(24)    & 0.3 &  0.247(35)     &  8.8 & 0.225(43) \\
      6.2            & 0.4 & 2.06(18)    & 0.8 &  0.364(99)     &  0.0 & 0.2628(11)\\
      \hline
   \end{tabular}
\end{table}

\begin{figure}[t]
\begin{center}
\epsfig{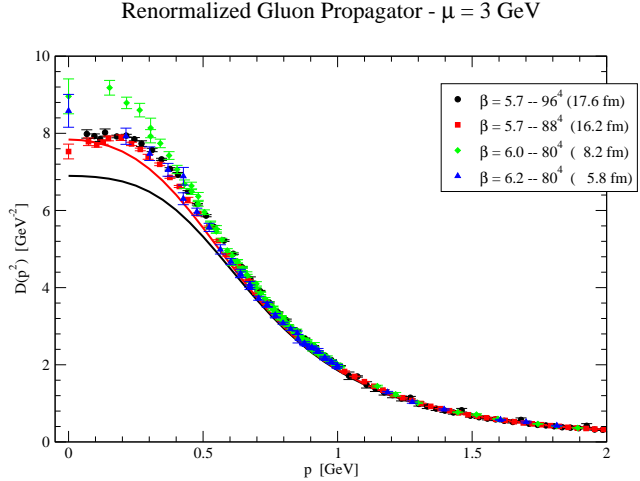}
\caption{Gluon propagator and the fit to (\ref{eq:Drgz}) using the linear extrapolated parameters. The red line
              was computed using the extrapolated $\beta = 6.2$ values, while the black line used the
              extrapolated $\beta = 6.0$ parameters.}
\label{fig:Drgz_ext}
\end{center}
\end{figure}

\noindent
and the corresponding zero momentum value is
\begin{equation}
   D(0) = \left\{ \begin{array}{lll}
   6.90(93) \mbox{ GeV}^{-2} & & \mbox{from the } \beta = 6.0 \mbox{ data set}, \\
      & & \\
   7.84(69) \mbox{ GeV}^{-2}   & & \mbox{from the } \beta = 6.2 \mbox{ data set}.
   \end{array} \right.
\end{equation}
In the computation of $D(0)$, given the poor quality of the linear extrapolation for
$\lambda^4$, we have used instead the fitted value from the largest physical volume.
Figure \ref{fig:Drgz_ext} shows the lattice gluon propagator for the largest physical volumes
computed using different lattice spacings and the extrapolated fits to (\ref{eq:Drgz}) as described above.

In the RGZ propagator, the parameter $m^2$ is related to the $\langle A^2 \rangle$. If one uses the figures
from the extrapolations it follows
\begin{equation}
   \langle g^2 A^2 \rangle_{{\tiny\, 3\,  \mathrm{GeV}}} =
    \left\{ \begin{array}{lll}
   2.71(41) \mbox{ GeV}^{2} & & \mbox{from the } \beta = 6.0 \mbox{ data set}, \\
      & & \\
   3.21(48) \mbox{ GeV}^{2}   & & \mbox{from the } \beta = 6.2 \mbox{ data set}.
   \end{array} \right.
\end{equation}
or
\begin{equation}
   \langle g^2 A^2 \rangle_{{\tiny\, 10\,  \mathrm{GeV}}} =
    \left\{ \begin{array}{lll}
   2.45(38) \mbox{ GeV}^{2} & & \mbox{from the } \beta = 6.0 \mbox{ data set}, \\
      & & \\
   2.90(43) \mbox{ GeV}^{2}   & & \mbox{from the } \beta = 6.2 \mbox{ data set}.
   \end{array} \right.
\end{equation}
The values are slightly below those reported in \cite{Dudal2010}.

\section{The Gluon Mass and Chiral Symmetry Breaking}

A gluon mass term in the QCD action is forbidden by gauge invariance and, therefore, a gluon mass $m_g$
has to be generated dynamically. A non-vanishing mass means that the gluon field is short ranged. 
We should be careful not to attribute a physical meaning to this ``massive'' gluon, given the already mentioned positivity 
violation, which is a indication of the unphysical (confined) nature of the gluon.
Besides providing the screening of the gluon, one may ask if there are
additional implications of having $m_g \ne 0$. In this section, we show that, within an effective field theory
for low energy QCD, a gluon mass is connected with chiral symmetry breaking, i.e. the theory either has
$m_g \ne 0$ and chiral symmetry is broken or chiral symmetry is restored and the gluon is a
long range field. This section is based in the work \cite{Oliveira2011a}.

In QCD the fundamental fields are associated with quarks and gluons. However, to describe the
low energy regime of QCD other fields can be included to define an effective theory.
Let us assume that the non-perturbative physics is mainly associated with the gluon sector.
Pure Yang-Mills theory has multi-gluon configurations as bound states. The simplest of these
bound states is a two gluon state. Given that the gluon belongs to the adjoint representation, the
two gluon state can be decomposed according to
$   8 \otimes 8 = 1 \oplus 8 \oplus 8 \oplus 10 \oplus \overline{10} \oplus 27 \, .$
The lowest dimensional irrep is a singlet and can be identified with glueball states. The
lightest glueball state has $J^{PC} = 0^{++}$ and a predicted mass of
$\sim 1.7$ GeV \cite{Chen2006}. Such a mass scale is well above the usual low energy mass scales,
$\sim 1$ GeV or lower, and therefore, from the point of view of an effective theory,
one expects the singlet to play a minor role. The next lower dimensional representations are
the two 8 representations. They distinguish amongst themselves because one of them is symmetric under interchange
of the gluons, while the other one is antisymmetric. Of the 8 irreps only the symmetric representation
can generate a scalar field, which can be written as
\begin{equation}
  \phi^a \propto d_{abc} F^b_{\mu\nu} F^{c \, \mu\nu} \, ,
\end{equation}
where $F^a_{\mu\nu}$ is the non-abelian Maxwell tensor. Of course, one can add to the above definition
a quark contribution given by, for example, $\overline q \, t^a q$, where $t^a$ are the generators of the
fundamental representation. Adding the two terms enables to estimate the contribution of quarks and
gluons to the effective field,
\begin{equation}
  \phi^a \approx \frac{ \langle F^2 \rangle}{\Lambda^3}Ê+ \frac{\langle \overline q \, q \rangle}{\Lambda^2} \, ,
\end{equation}
where $\Lambda \sim \Lambda_{QCD}$ is a non-perturbative mass scale. Plugging into this the gluon condensate
$\alpha_s \langle F^2 \rangle = 0.04 $ GeV$^4$ and the light quark condensate
$\langle \overline q \, q \rangle = ( - 270 \mbox{ MeV} )^3$, it follows that the ratio gluon to quark content
of $\phi^a$ is around 7.

Let us consider an effective theory which includes the gluon field $A_\mu$, the quark fields $q_f$ , where $f$ is a flavor
index, and an effective scalar field $\phi^a$ that belongs to the adjoint representation of the SU(3) color group.
In the following we will assume that the non-perturbative physics is contained in $\phi^a$. Furthermore,
being an effective field theory, it should describe hadronic physics only in the low energy regime and
it does not need to be renormalizable. The effective Lagrangian reads
\begin{eqnarray}
 \mathcal{L} & = & - \frac{1}{4} F^a_{\mu\nu} F^{a \, \mu\nu} +
                                 \sum_f \overline q_g \left\{ i \gamma^\mu D_\mu - m_f \right\} q_f \nonumber \\
   &   & +  \, \frac{1}{2}ÊD^\mu \phi^a \, D_\mu \phi^a  -  V_{oct} ( \phi^a \phi^a )
            + \mathcal{L}_{GF} + \mathcal{L}_{ghost}  \nonumber \\
   &    &  - \, G_4 \sum_f \left[Ê\overline q _f \, t^a \, q \right] \, \phi^a \nonumber \\
   &   &  - \, G_5 \sum_f \left[Ê\overline q _f \, q \right] \, \phi^a  \phi^a
             - \, F_1 \sum_f \left[Ê\overline q _f \, q \right] \, d_{abc} \phi^b  \phi^c \nonumber \\
   &   &   - \, F_2 \sum_f \left[Ê\overline q _f \, t^a \gamma^\mu q \right] \, D_\mu \phi^a
              - \, F_3 \sum_f \left[Ê\overline q _f \, t^a \gamma^\mu D_\mu q \right] \, \phi^a  + h.c.
\end{eqnarray}
where $D_\mu = \partial_\mu + i g T^a A^\mu$ is the covariant derivative, $T^a$ the SU(3)
generators, $m_f$ the current quark mass associated with flavor $f$,
$V_{oct}$ the effective potential associated with the scalar field.
$\mathcal{L}_{GF}$ is the gauge fixing part of the Lagrangian and
$\mathcal{L}_{ghost}$ contains the ghost terms.
The Lagrangian is gauge invariant, excepts for the $\mathcal{L}_{GF}$ term.
The effective gauge coupling constant $g$ parameterizes residual interactions and it should be a small number,
i.e. one expects the theory can be treated perturbatively.
The new interactions with the scalar field, the terms proportional to $G_4$, $G_5$, $F_1$, $F_2$ and $F_3$,
where written assuming flavor independence of strong interactions.

$\mathcal{L}$ includes the QCD Lagrangian and verifies the usual soft-pion theorems of
chiral symmetry at low energy. The new interactions introduce new vertices, not present in the original QCD
Lagrangian, which contribute to quark processes. Note that the only new quark color singlet operator mimics the
$^3P_0$ model describing OZI-allowed mesonic strong decays.

The $\phi^a$ kinetic term couples to a quadratic gluon term through the operator
\begin{equation}
 \frac{1}{2} g^2 \phi^c ( T^a T^b )_{cd} \phi^d A^a_\mu A^{b\,\mu} .	
\end{equation}
If the scalar fields acquires a vacuum expectation value without breaking color symmetry, i.e.
\begin{equation}
  \langle \phi^a \rangle = 0 \quad \mbox{ and } \quad \langle \phi^a \phi^b \rangle = v^2 \, \delta^{ab} ,
\end{equation}
given that for the adjoint representation $ tr ( T^a T^b ) = N_c \delta^{ab}$, the gluon mass reads
\begin{equation}
  m^2_g = N_c g^2 v^2 ,
\end{equation}
where $N_c = 3$. From the definition it follows that $\langle \phi^a \phi^b\rangle$, i.e. $v^2$,
and therefore the gluon mass is gauge invariant.
The proof of gauge invariance follows from the transformation properties of $\phi^a$.

In the same way, the operator $G_5 \, \left[ \overline q  \, q \right] \, \phi^a \phi^a$ shifts the quark masses
giving rise to a constituent quark mass
\begin{equation}
  M_f = m_f - (N^2_c - 1) \, G_5 \, v^2 = m_f - \frac{N^2_c - 1}{N_c} \, \frac{G_5}{g^2} \, m^2_g \, .
 \label{Mf_mg}
\end{equation}
For light quarks, the constituent quark mass is given by the quark self energy which, in the model,
is linked with the gluon mass.
 Note, for our definitions, that $G5$ is a negative number.
If the constituent mass for the light quarks vanishes, chiral symmetry 
should be broken dynamically, whereby the relation
$M_f \propto m^2_g$ in the effective model links chiral symmetry with a finite effective gluon mass. 

The quark condensate $\langle \overline q \, q \rangle$, an order parameter for chiral symmetry breaking,
can be computed in the model as a function of the constituent quark mass, the gluon mass and the
theory cut-off - see \cite{Oliveira2011a} for details. Then, if one identifies the gluon mass with the
mass measured from the lattice using a simple pole propagator, $m_g = 634$ MeV, together with
$M_f = 330$ MeV and $\langle \overline q \, q \rangle = ( - 270 \mbox{ MeV})^3$, one is able
to estimate some of the theory parameters:
\begin{displaymath}
\overline\omega = 879 \mbox{ MeV}, \quad
gv = 366 \mbox{ MeV} \quad
\mbox{ and } \quad
\frac{G_5}{g^2} = -0.31 \mbox{ GeV}^{-1} \, ,
\end{displaymath}
where $\overline\omega$ is the theory's cut-off.

\section{Testing a Model Prediction}

The effective model relates the constituent quark mass $M$ and the gluon mass $m_g$
through equation (\ref{Mf_mg}). For a vanishing current mass, equation (\ref{Mf_mg}) predicts a
constant value for the ratio $M/m^2_g$, at least at tree level. 
This result can be tested looking at the solutions of the
Schwinger-Dyson equations. In the following we will use the results published in \cite{Aguilar2011}.
For the gluon and ghost propagators, the authors used the results of lattice QCD simulations and solved the gap equation for a massless fermion. The calculation does not take into account fermion loops and can be viewed as a quenched approximation.

\begin{figure}[t]
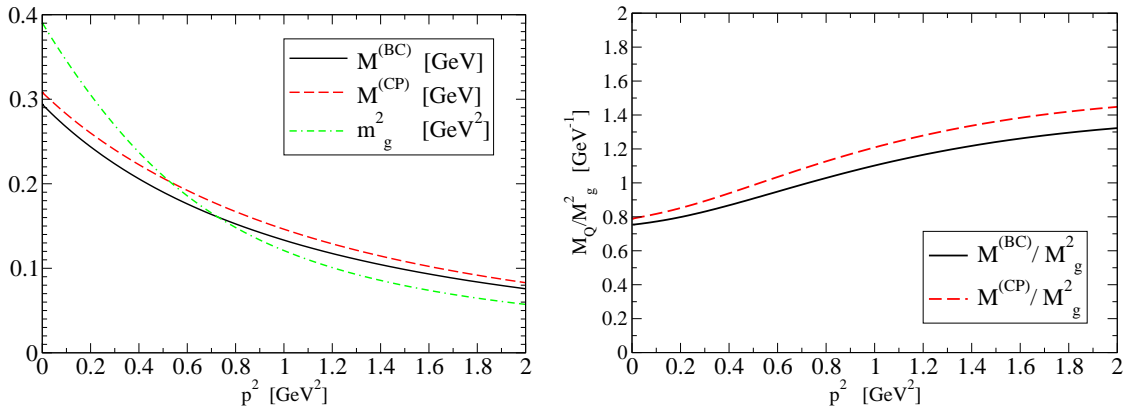

\begin{center}
\epsfig{file=masses_q_g.eps,height=2.1in} \,\,
\epsfig{file=mq_over_m2glue_mod.eps,height=2.1in}
\caption{On the left hand side, the plot shows the quark masses from solving the fermionic SDE gap equation, using different ans\"atze for the quark-gluon vertex, and the squared gluon mass computed from quenched lattice simulations. Note that $M(p^2$) depends slightly on the definition of the quark-gluon vertex. On the right hand side, the plot shows the ratio $M/m^2_g$.}
\label{fig:SDE}
\end{center}
\end{figure}

In \cite{Aguilar2011}, the fermionic gap equation was solved for two different ans\"atze for the quark-gluon vertex,
a non-Abelian improved version of the Ball-Chiu vertex and an improved version of the Curtis-Pennington vertex.
The choice of vertex leads to slightly different quark mass. In order to distinguish, the results of the Ball-Chiu vertex
will be referred as ÓBCÓ, while the results from using the Curtis-Pennington vertex will be referred as CP. Figure
\ref{fig:SDE} shows $M$ computed from the Schwinger-Dyson equations for the different vertex ans\"atze, together
with $m^2_g$, as a function of $p^2$ and, on the right hand side, the ratio $M/m^2_g$. The plots shows that
$M/m^2_g$ increases slightly.
If one looks at the maximal momentum range where the lattice gluon propagator can be fitted by a simple pole, i.e.
if one compares the ratios up to momenta $p  \sim 0.5$ GeV, then
$M/m^2_g$ changes by less than 8\%, relative to its zero momentum value, when using
the BC quark-gluon vertex and less than 10\% when using the CP vertex.

\section{Results and Conclusions}

We have currently a fair description of the gluon propagator over all momentum ranges. To extract
the various parameters modeling the propagator, it would be desirable to perform a high statistic and
large volume simulation. 

The results of lattice simulations and Schwinger-Dyson equations show that the gluon propagator behaves
as a dynamically massive gauge boson in the infrared region, see also the discussion in
\cite{Pennington2011} and references therein, 
 and, the effective model sketched here, shows a 
connection between
the gluon mass and chiral symmetry breaking. Comparing the tree level mass ratio prediction with the solutions
of the Schwinger-Dyson equations we found good agreement in the low energy regime.

\bigskip
The authors acknowledge financial support from the Brazilian
agencies FAPESP (Funda\c c\~ao de Amparo \`a Pesquisa do Estado de
S\~ao Paulo), CNPq (Conselho Nacional de Desenvolvimento
Cient\'ifico e Tecnol\'ogico) and Research-Foundation Flanders (FWO Vlaanderen).
 OO acknowledges financial support from FCT under contract PTDC/\-FIS/100968/2008.

\end{document}